Dissociative Excitation of Acetylene Induced by Electron Impact

Electron Impact Excitation of Acetylene

Juraj Országh*, Marián Danko, Peter Čechvala, Štefan Matejčík

*Department of Experimental Physics, Faculty of Mathematics, Physics and Informatics, Comenius University in Bratislava, Mlynská dolina F-2, 842 48 Bratislava, Slovakia*

*juraj.orszagh@uniba.sk

Abstract

The optical emission spectrum of acetylene excited by monoenergetic electrons was studied in the range 190-660 nm. The dissociative excitation and dissociative ionisation associated with excitation of the ions initiated by electron impact were dominant processes contributing to the spectrum. The spectrum was dominated by the atomic lines (hydrogen Balmer series, carbon) and molecular bands (CH(A-X), CH(B-X), $CH^+$(B-A), and $C_2$). Besides the discrete transitions we have detected the continuum emission radiation of ethynyl radical $C_2H$(A-X). For most important lines and bands of the spectrum we have measured absolute excitation-emission cross sections and determined the energy thresholds of the particular dissociative channels.

## Introduction

The collisions of electrons with molecules play an important role in many environments, e.g., in planetary atmospheres, different plasmas, radiation chemistry, etc. Electron induced fluorescence (EIF) spectroscopy is one of the methods that can be applied to study the inelastic collisions of electrons with molecules and provide kinetic data on the excitation, dissociative excitation and ionization processes. Moreover, it is a suitable method to study the emission spectra of the atoms and molecules. One of the main advantages of the method is the possibility to excite even states that cannot be induced by photon excitation due to optically forbidden transitions and that excitation occurs at well-defined energy.

Acetylene is the smallest hydrocarbon with triple bond between carbon atoms. From the spectroscopic point of view acetylene is interesting for its abundance in the space (Boyé et al. 2002, Tanabashi et al. 2007, Lacy et al. 1989, Brook et al. 1996, Tucker et al. 1974). The molecule has also relevance to different plasma technologies (Childs et al. 1992, Smith et al. 2001, Somé et al. 1995). As a small hydrocarbon its presence can be expected also in the boundary region of the plasma in thermonuclear reactors. The sources of impurities like carbon or CH were localized for example in tokamak TEXTOR (Pospieszczyk et al. 1987). Relatively strong emission of carbon was detected in the fusion reactor plasma by means of optical emission spectroscopy (OES) (Zimin et al. 2015) and the chemical erosion of carbon in the tokamak was specifically addressed using OES as well (Brezinsek et al. 2007). Extensive database of reactions of small hydrocarbons with plasma electrons and protons relevant especially for thermonuclear fusion edge plasmas were published by Janev and Reiter (Janev et al. 2002, Janev et al. 2004). Ethynyl radical ($C_2H$) which can be formed from acetylene by photon or electron impact belongs also to the most abundant interstellar polyatomic molecules (Tucker et al. 1974). According to (Beuther et al. 2007) it is present from the onset of massive star formation but later it is transformed into other compounds. In the outer cloud regions its abundance remains high due to constant replenishment of carbon from photon induced dissociation of CO (Beuther et al. 2007).

Acetylen presence was confirmed by the Cassini Ultraviolet Imaging Spectrometer in the atmosphere of Titan at altitudes between 450km and 1400km with highest abundance at 450km (Shemansky et al. 2005, Liang et al. 2007). The solar Lyman $\alpha$ radiation is absorbed mostly in the agnostosphere of Titan where it acts as a primary initiator of photochemical processes and where many hydrocarbon compounds are created (Koskinen 2011). The generated photoelectrons are also an efficient driver of chemical processes (Krasnopolsky 2009). These reactions produce stratospheric haze seed particles such as $C_2H$. In such environment, the data on electron-molecule processes are relevant. The presence of acetylene in the atmospheres of some exoplanets was also suggested by Oppenheimer et al. (2013). It is also referred as inverse biosignature gas in case of Titan or exoplanets (Seager et al. 2012; Seager et al. 2013). Its importance is not given only by its abundance in space but also by the fact that it provides a link in the chemistry of acetylene and higher order polyacetylenes.

The information about the acetylene and ethynyl radical in space are usually obtained from the absorption spectra in the infrared spectral range (vibrational states) or in the microwave range (rotational states). The **EIF** provides information about excited states of the molecules and about radiative electronic transitions of the molecules. This information can be potentially useful for the absorption studies in the UV-VIS range of the spectra. EIF method allows determining also the kinetic parameters of the electron induced processes, which are important for the simulation of the processes in such media as planetary atmosphere, nebulae, or plasmas.

The emission spectrum of acetylene in the UV-VIS range exhibits several atomic lines and molecular bands. Majority of the processes responsible for emission of photons originate from dissociative excitation processes. The acetylene molecules excited to higher excited states dissociate into excited fragments (molecular or atomic) and one or more neutral fragments. Thus, the emission of the H, C, $C^+$, various CH, $CH^+$, $C_2$ states and the continuum emission of the $C_2H$ was detected in the emission spectrum of acetylene in the UV–VIS range (Beenaker & de Heer 1974, Okabe 1975).

In 1974 (Beenaker & de Heer 1974) studied dissociative excitation of acetylene by electron impact. They focused on selected lines and bands of the excited fragments CH, H, C and $C_2$ within the electron energy range from **thresholds up to 1000 eV** and determined as well the threshold energy values. The emission spectrum was studied within the range 185-900 nm. Later (Beenaker et al. 1975) published a study of vibrational and rotational structure of CH($A^2\Delta$ - $X^2\Pi$) transition. The spectrum of this transition was found to be independent of the electron impact energy within the range from 15-1000 eV and was composed of the overlapping vibrational transitions 0-0, 1-1 and 2-2. The emission spectrum of $C_2H_2$ at **200 eV** and cross section for $CH^+$($B^1\Delta$ – $A^1\Pi$) transition were determined by (Tsuji et al. 1975). They determined the threshold of $CH^+$($B^1\Delta$ – $A^1\Pi$) and identified several other molecular bands. Absolute emission cross sections of $H_\alpha$(3–2) and CH ($A^2\Delta$-$X^2\Pi$) were object of the study by (Möhlmann et al. & de Heer 1977). They were determined in the energy region of 0-2000 eV together with several other hydrocarbons such as methane.

The fluorescence spectrum of $C_2H_2$ was studied also in association with photon excitation. Okabe (1975) suggested that the dissociative excitation processes into $C_2H$ fragment is responsible for the broad continuum detected in the acetylene spectra. This continuum has maximum approximately at 480 nm. In the experiment the authors used a krypton lamp radiating at 116.5 nm and 123.6 nm for excitation of acetylene. The energetic threshold for the continuum acetylene radiation was found to be 9.5±0.2 eV. There are several other papers (Boyé et al. 2002, McDonald et al. 1978, Schmieder 1982, Saito et al. 1984, Hsu et al. 1992) reporting experimental excitations of acetylene by photons observing radiation mostly from $C_2H$, $C_2$, etc.

It is possible to conclude that the acetylene emission spectrum is relatively well described. On the other hand, there are still areas where more information is needed. In this paper we will describe the continuum radiation of the ethynyl fragment which has not been studied in electron induced processes. In addition, we are going to present absolute emission cross sections and determinate the threshold energies corresponding to dissociation processes including Balmer hydrogen emission, and excited states of CH, $CH^+$ and $C_2$.

Experimental setup

Acetylene excitation by electron impact was studied using the experimental apparatus utilizing perpendicular beams of electrons and molecules shown in the Fig. 1. and described in detail in (Danko et al. 2013).

The electron beam has energetic resolution of **300-400 meV** full width at half maximum (FWHM) with electron current approximately 750 – 900 nA and is generated by trochoidal electron monochromator (TEM) (Matuska et al. 2009) which utilizes crossed electric and magnetic fields in the dispersion region. The beam resolution was determined from the shape of the measured excitation-emission cross sections of Helium He I (1s2p - 1s4d) transition (NIST). The fluctuations of the electron current were below 5% in case of spectral measurements and less than 10% with changing the acceleration voltage. **The electron and molecular beam cross each other at the angle of 90°**. As the experiment was performed in regime of binary collisions i.e. only one electron collides with only one molecule and the number of molecules per second in the molecular beam is constant and independent on the electron accelerating voltage the fluctuations of the electron current should translate into the fluctuations of the measured photon yield within 5% for spectral measurements and within 10% for Photon Efficiency Curves (PEC) measurements.

The purity of the used gas was 99.6% and it was introduced into the reaction chamber from effusive capillary source regulated by leak valve (see Danko et al. 2013 for details). The working pressure in the vacuum chamber was $1x10^{-4}$ mbar and the background pressure in the system without gas was less than $5x10^{-8}$ mbar.

The light emission occurring during the deexcitation was collected in 90° angle in reference to the electron and molecular beam and within the 26° acceptance angle (see Figure 1). Spherical mirror with the focus in the centre of reaction volume positioned opposite the collection lens (L1 in the Figure 1) was used to increase the photon collection efficiency. No method for determination of polarisation of the emitted light was used in this experiment yielding error less than 5%. The slits width of the optical monochromator was set to 500 μm (corresponding to spectral resolution 2 nm) in order to achieve reasonable compromise between sensitivity and optical resolution when needed. Higher resolution spectra were measured with 100 μm slits (corresponding to spectral resolution 0.4 nm). The light was detected by Hamamatsu H4220P photomultiplier working in the photon counting regime. The detector is sensitive in the spectral region between 185 nm and 710 nm. The uncertainty of the spectral calibration is between 20% for shortest and longest wavelength in the range and approximately 5% for the middle section of the region.

The cross section of the reaction volume from the lens side of view is approximately 0.5x1 mm and it is given by the cross section of the electron and molecular beams. The instrumental field of view is given by the optical monochromator acceptance angle and the parameters and positions of the used lenses and mirror. The width of the spot from which the emitted photons can reach the photomultiplier is approximately 3 mm for 100 μm slit width.

The black-body radiation of the hot filament used as a source of electrons was supressed by special mounting and most of the inner surfaces of the vacuum system were covered by colloidal graphite to diminish possible light reflections. To further reduce the negative effects of the background filament radiation its spectrum was recorded under high vacuum conditions ($5x10^{-8}$ mbar) and subtracted from measured molecular spectra.

The electron energy was calibrated by measurement of the cross section of the (0,0) band of $N_2$ ($C^3\Pi_u$ – $B^3\Pi_g$) at 337 nm (Országh et al. 2012, Zubek 1994) and He I (1s2p $^3P^0_{1,2}$ - 1s4d $^3D_{1,2,3}$) 447.14 nm emission line (NIST). The $N_2$ ($C^3\Pi_u$ – $B^3\Pi_g$)(0,0) cross section exhibits a pronounced peak at 14.1 eV. The cross section for the He I (1s2p $^3P^0_{1,2}$-1s4d $^3D_{1,2,3}$) transition exhibits a step at 23.736 eV. The electron energy calibration was performed in the mixture of $C_2H_2$:$N_2$:He 1:1:1 (established by absolute pressure gauge) to avoid problems with different charging effects of the electron monochromator electrodes in different gases. The difference between energy scales calibrated by nitrogen and helium was less than 0.015 eV.

Two different measurement modes were used in the present study, (i) fluorescence spectra induced by monoenergetic electrons at fixed electron energy and (ii) Photon Efficiency Curves (PEC) – the intensity of photons of specific wavelength detected as a function of the electron energy, sometimes referred to as relative excitation function. PECs are proportional to dissociative excitation-emission cross sections (DEE CS) for and can be calibrated to its absolute values. Additionally, PEC curves can be used for determination of the apparent thresholds in PEC which are related to the thresholds for the dissociative excitation reactions (Stano et al. 2001).

## Results and discussion

### *Spectrum*

We have recorded the electron induced UV-VIS spectrum of acetylene in the spectral range from 190 nm to 700 nm. The spectral range 190-257 nm was recorded at electron energy of 70 eV and low resolution of the optical monochromator (2 nm FWHM) and the rest at 50 eV and higher resolution of the optical monochromator (0.4 nm FWHM). The spectral range 305-530 nm is presented in the Figures 2a and 2b. All the spectra presented in this paper were corrected for the spectral response of the apparatus. The intensities in the spectra were calibrated to absolute values according to the known emission cross sections for the dissociative excitation-emission channel that we will give more details on later. The spectra exhibit lines and bands corresponding to dissociative excitation of acetylene by electron impact (Beenaker et al. 1975, Tsuji et al. 1975).

Due to lower sensitivity of the optical system in the region 190-257 nm it was necessary to decrease the resolution of the optical monochromator (using 500 μm for both slits). We were able to identify two C I lines at 193.1 nm (C I (**$2s^22p3s$ $^1P^0$** – **$2s^22p^2$ $^1D$**)) and 248.4 nm (C I (**$2s^22p3s$ $^1P^0$** – **$2s^22p^2$ $^1S$**)) and the Mulliken system $C_2$ ($D^1\Sigma_u^+$ - $X^1\Sigma_g^+$) according to (Möhlmann & de Heer 1975, Schmieder 1982 and NIST). The C I (**$2s^22p3s$ $^1P^0$** – **$2s^22p^2$ $^1D$**) line at 193.1 nm was most probably blended with the C II (**$2s^24p$ $^2P^0$** – **$2s2p^2$ $^2P$**) line at 192.8 nm (Pang et al. 1987).

In the spectral region 305-530 nm we have recorded the spectrum with improved spectral resolution at 50 eV electron energy (Fig. 2a and 2b). The absolute intensity calibration of this spectral region was based on the emission cross section value for $H_\beta$ (4-2) published in (Beenaker et al. 1974). The authors estimated the error of their cross section values to be between 22% at short wavelengths (**~190 nm**) and 8% (**~650 nm**). As there is no specific error given for the $H_\beta$ cross section let us assume linear dependence of the error on the wavelength. In such case the error for the $H_\beta$ cross section would be approximately 13% and contributes to our cross section values. In the Figure 2a we have recognised several structures, which were identified on the basis of data from (Beenaker et al. 1975, Tsuji et al. 1975, Schmieder 1982 and Craig et al. 1982). The spectrum in Fig. 2a is dominated by the molecular emission bands CH ($B^2\Sigma^-$ - $X^2\Pi$) (0,0), CH ($C^2\Sigma^-$ - $X^2\Pi$) (1,0) and the ion system $CH^+$($B^1\Delta$ - $A^1\Pi$) (0,0). We have detected $C_2$ ($C^1\Pi_g$ - $A^1\Pi_u$) or so-called Deslandres d'Azambuja system as well (Tsuji et al. 1975, McDonald et al. 1978, Schmieder 1982 and Acquaviva et al 2002). Additionally, we can see the Balmer series lines $H_\delta$ (6 - 2) – $H_\eta$ (9 - 2) and possibly C II ($2s^24s$ $^2S$ - $2s^23p$ $^2P^0$) line (391 nm) (NIST).

The Figure 2b shows the 420-660 nm range of the spectrum. We are able to identify CH($A^2\Delta$ - $X^2\Pi$) and $CH^+$($B^1\Delta$ - $A^1\Pi$) (blended with the triplet $CH^+$(b - a) system) bands (Furuya et al. 1997). In this range $C_2$ ($d^3\Pi_g$ - $a^3\Pi_u$) or Swan system (Tanabashi et al.

2007, McDonald et al. 1978, Parriger et al. 1994) was detected as well (two sequences Δυ = +1, 0). In the region 530-570 nm also the Δυ = -1 sequence of vibrational Swan system was detected, however, we do not present this part of spectrum as the signal-to-noise ratio was very small in this region and the the Δυ = -1 sequence was barely recognizable. The positions of these peaks correspond to Parriger et al. 1994. The relative intensities of the Δυ = +1 and Δυ = 0 transitions are similar to those measured by Parriger et al. 1994 from laser induced plasma. Due to relatively low signal-to-noise ratios in our experiment the shapes of the bands are slightly deformed in comparison to Parriger et al. 1994 especially for Δυ = 0 but the repeated measurement with 500μm slit width at the expense of the optical resolution confirmed the shape is similar.

*Continuum $C_2H$* ($\tilde{A}^2\Pi$ - $\tilde{X}^2\Sigma^+$)

It may not be evident from Figure 2a and 2b but a large part of the measured spectrum is superimposed on a continuum radiation of ethynyl $C_2H$ ($\tilde{A}^2\Pi$ – $\tilde{X}^2\Sigma^+$) fragment. Due to the fact that the threshold energy for dissociation of acetylene into $C_2H(\tilde{A}^2\Pi)$ is lower than the thresholds for formation of other excited fragments in the studied spectral range, the spectrum recorded at 13 eV exhibits only the continuum radiation $C_2H$ ($\tilde{A}^2\Pi$ – $\tilde{X}^2\Sigma^+$). According to (Boyé et al. 2002 and Saito et al. 1984) this continuum radiation covers range from 400 nm to infrared region.

We have recorded it in the 400-650 nm range and it is shown in the Figure 3. For comparison we have digitised three spectra $C_2H$ ($\tilde{A}^2\Pi$ - $\tilde{X}^2\Sigma^+$) from (Boyé et al. 2002) obtained in photolysis of acetylene. The energies/wavelengths of the VUV photons were selected to excite the acetylene to different Rydberg states, which predissociated into excited ethynyl fragment. In the spectral region 400-460 nm present spectrum resembles the 11 eV photolysis spectrum (excitation of Rydberg states $6d\pi^1\Sigma_u^+ + 6d\sigma^1\Pi_u$). On the other hand, the spectral region above 520 nm is more similar to the excitation of lower Rydberg states $\tilde{G}(4s\sigma)^1\Pi_u$ and $\tilde{H}(3d\delta)^1\Pi_u$ (9.98 eV line). Based on this observation it can be suggested that the 13 eV electrons excite several Rydberg states so that the resulting continuum spectrum of $C_2H$ measured at 13 eV is superposition of spectral dissociation of mixture of these states.

*Dissociative excitation-emission cross sections and threshold energies*

Most of the photon efficiency curves (PECs) measured at particular wavelength (associated with main transition) were blended with other transitions. As an example we present two PECs recorded at the wavelength of the Balmer β and γ (H(4 - 2), respectively H(5 - 2)) (Figure 4.). In the Figure 4 we can see that the H(4 - 2) radiation was blended by the ethynyl continuum radiation. The H(5 - 2) line was blended both by ethynyl continuum radiation and CH($A^2\Delta$ - $X^2\Pi$) radiation. One way to deal with the problem of more mixed lines or bands is to use data analysis and remove the superposed transitions. After removal of the blended lines or bands by subtraction of corresponding contributions, we have obtained "pure" Balmer emission lines of acetylene. **The Balmer PECs are affected by the long lifetime of the excited states and finite instrumental field of view which can lead to signal losses. The amount of lost signal was estimated in Tables 1a and 1b for specific conditions.** We have measured the PECs for all strong emission lines present in the emission spectrum (Figure 2a and 2b) and also in the unpublished parts of the spectrum. The PECs were evaluated for the apparent threshold and cusps, which we related to the thresholds for particular dissociative excitation channels.

We have determined the experimental values of energy thresholds for dissociative processes associated with detected radiation and compared them with values determined using thermodynamic calculation. We calculated the theoretical threshold values from the reaction enthalpies for the dissociation into ground state and the excitation energies of the upper states, using NIST Chemistry WebBook database (NIST). The calculations were done for several dissociative excitation channels, and experimental values were associated with the ones they could correspond with. The values for Balmer H (n - 2) transitions, $C_2$ and CH bands, C lines, and $C_2H$ band are shown in the Tables 2 and 3. The experimental values are usually higher than the calculated theoretical thresholds. **There are several reasons for this behaviour such as noise in the data which affects fitting errors, electron energy distribution of our experimental device, and the energy distribution between the dissociative products and internal energy of dissociative products**. We have estimated the error bars of obtained experimental thresholds to be 0.5 eV and in case of very noisy data to 1 eV. Not all of the possible dissociative excitation channels of acetylene molecule were experimentally observed as active. Some of the estimated thresholds were lower than the lowest calculated threshold of the certain process, which means that it is associated with a different process.

The photon efficiency curves (PEC) were calibrated to the absolute values in order to obtain the dissociative excitation-emission cross sections. The calibration was based on the absolute values of the emission spectra at 50 eV. The spectrum was corrected for the apparatus sensitivity so the relative intensities of the observed bands and lines were correct. In the next step the intensity of the spectrum was normalized to the cross section value of the transition $H_\beta$ (4 – 2) at 50 eV published by (Beenaker & de Heer 1974). Finally, the relative PECs were calibrated to the values of the corresponding bands of the spectrum at 50 eV.

The cross sections for the dissociative excitation and emission of the hydrogen Balmer lines upon electron impact are shown in the Figure 5a. The cross sections for H(4 - 2) line agree within the error bars with values of (Beenaker & de Heer 1974) at 60 eV and 40 eV. The cross section values for given electron energy are decreasing with increasing excitation level n. Similar behaviour was observed for dissociative excitation of methane (Danko et al 2013). We were able to determine the cross section for H transition (5 - 2) for the first time. Unfortunately, the shape of the PEC of $H_{\varepsilon}$(7 - 2) could not be determined correctly as the fluorescence signal at the wavelength 397.5 nm was strongly superimposed with several transitions that we were unable to identify and subtract from the PEC of $H_{\varepsilon}$(7 - 2). The list of the values of the cross sections at selected electron energies is presented in the Table 4. Taking into account the uncertainties mentioned earlier the overall error of the cross section values is ranging from 18% to 26%.

The finite field of view of the experimental apparatus has a negative impact on the measurement error as the species with the longer radiative lifetimes can escape the volume from which the photons can be detected before the emission occurs. Let us assume the excited Hydrogen atom is generated in the middle of reaction volume, that it can move to any direction with the same probability and that the photon can be emitted into any direction with the same probability. Then due to the finite field of view of the experimental device significant portion of photons generated from H (n-2) where n > 5 will not reach the photomultiplier. The data for such situation are estimated in the following table for excited species temperatures 300K and 500K. The apparent PECs for the H (n-2) where n > 5 transitions are shown in the figure 5b and due to the mentioned incomplete detection problem they were not calibrated to absolute cross section values.

Tab. 1a. Estimate of the photon loss due to the finite field of view of the experimental apparatus and long lifetimes of excited species at H* temperatures of 300K and 500K.

| Transition | H* lifetime (ns) | Distance travelled per lifetime at 300K (mm) | Distance travelled per lifetime at 500K (mm) | Photons lost at 300K (%) | Photons lost at 500K (%) |
|---|---|---|---|---|---|
| H (3 - 2) | 22.7 | 0.051 | 0.065 | 0 | 0 |
| H (4 - 2) | 119 | 0.265 | 0.343 | 0 | 0 |
| H (5 - 2) | 395 | 0.883 | 1.140 | 0 | 0 |
| H (6 - 2) | 1030 | 2.300 | 2.960 | 54.8 | 66.2 |
| H (7 - 2) | 2280 | 5.090 | 6.570 | 80.9 | 85.3 |
| H (8 - 2) | 4520 | 10.100 | 13.000 | 90.5 | 92.6 |

Let us now try to estimate the velocity of the excited hydrogen atoms in the experiment. Let us assume that at the cross section threshold all the kinetic energy of the electron is transferred to the molecule and it is used for dissociation of the molecule into the products, electronic excitation of the products and the excess energy is transformed into the kinetic energy of the products. For the sake of estimation let us also assume that none of the energy is transformed into the rotational and vibrational excitation of the products. In such case the kinetic energy will be distributed among the products according to their momentum. Under these circumstances the kinetic energy of the products would be given by the difference between the calculated threshold energy and measured threshold energy in Table 2. For such case the kinetic energies of the excited H atoms, the distances travelled per their lifetime and loss of photons due to limited field of view of the apparatus are shown in the table 1b. The estimate is calculated for the first threshold in each of Balmer series PECs i.e. for the process $e+C_2H_2 \rightarrow e+C_2H+H^*$ apart from the case of H (7 – 2) where the measured experimental threshold value is slightly lower then theoretical one based on enthalpy of formation and the estimate cannot be calculated.

Tab. 1b. Photon loss due to the finite field of view of the experimental apparatus based on H* kinetic energy estimated from the difference of the measured and theoretical thresholds of the processes.

| Transition | H* lifetime (ns) | Estimated kinetic energy (eV) | Distance travelled per lifetime (mm) | Photons lost (%) |
|---|---|---|---|---|
| H (3 – 2) | 22.7 | 3.11 | 0.556 | 0 |
| H (4 – 2) | 119 | 1.32 | 1.900 | 42.1 |
| H (5 – 2) | 395 | 1.99 | 7.74 | 87.6 |
| H (6 – 2) | 1030 | 3.75 | 27.7 | 96.6 |
| H (7 – 2) | 2280 | N/A | N/A | N/A |
| H (8 – 2) | 4520 | 2.73 | 10.4 | 90.8 |

The percentage of photon loss is extremely high especially for H (n – 2) where n > 4. **It is probable that such an estimate is strongly affected by neglecting the amount of energy transformed into the rotational and vibrational excitation of the products, noise in the measured signal and energetic resolution of the electron beam, so it is probable that data shown in Table 1a are more realistic.** The cascade excitations from higher levels is another unknown factor affecting the calculation.

Beside the cross sections for the Balmer lines, we were able to determine cross sections for dissociative excitation and emission for two CI lines (Figure 6a), for selected $C_2$ (Figure 6b) and CH (Figure 6c) bands. The PEC of $CH^+$ ($B^1\Delta$ - $A^1\Pi$) transition was recorded at 350.7 nm. There was no $C_2H$ continuum emission at this wavelength, but according to the threshold values determined at measured emission function it was possible to identify admixture of the nitrogen second positive system transition (2,3) (nitrogen was present in trace amounts in the sample) and $C_2$ Deslandres d'Azambuja transition. After subtraction of these two parasitic emissions we have obtained the cross section for dissociative excitation and emission for this product. From the PEC we were able to determine specific processes leading to formation of $CH^+$(A) ion (see Table 3).

The $C_2H$ ($A^2\Pi$ - $X^2\Sigma^+$) continuum radiation is induced by electron impact. This radiation is overlapped with several atomic lines and molecular bands. For this reason, we have to find a wavelength, where the continuum is free of the atomic lines and molecular bands. The most suitable we have found the 526 nm wavelength, where the PEC was measured. The cross section for dissociative excitation and emission of $C_2H$ ($A^2\Pi$ - $X^2\Sigma^+$) at 526 nm is presented in the Figure 7.

Tab. 2. Calculated (based on enthalpy of formation) and experimentally determined threshold energies for hydrogen and carbon atoms and their comparison with previously published values. Dissociative channels for thresholds are proposed.

| Transition | Wavelength [nm] | Reaction | Calculated threshold energy [eV] | Measured threshold energy [eV] | Published value [eV] |
|---|---|---|---|---|---|
| $H_\alpha$ (3 – 2) | 657 | $e+C_2H_2 \rightarrow e+C_2H(X)+H^*$ | 16.96 | 20.2±0.5 | |
| | | $e+C_2H_2 \rightarrow e+C_2H(B)+H^*$ | 21.82 | | |
| | | $e+C_2H_2 \rightarrow e+C+CH+H^*$ | 25.61 | 27.4±0.5 | |
| | | $e+C_2H_2 \rightarrow e+C_2(C)+H+H^*$ | 27.33 | | |
| | | $e+C_2H_2 \rightarrow e+2C+H+H^*$ | 29.13 | 30.9±0.5 | |
| | | $e+C_2H_2 \rightarrow e+C+CH(C)+H^*$ | 29.85 | | |
| $H_\beta$ (4 – 2) | 486.7 | $e+C_2H_2 \rightarrow e+C_2H(X)+H^*$ | 17.63 | 19.0±0.5 | 20.6±1 (Beenaker & de Heer 1974) |
| | | $e+C_2H_2 \rightarrow e+C_2H(A)+H^*$ | 18.08 | | |
| | | $e+C_2H_2 \rightarrow e+C_2H(B)+H^*$ | 22.48 | 23.0±0.5 | |
| | | $e+C_2H_2 \rightarrow e+C_2+H+H^*$ | 23.62 | | |
| | | $e+C_2H_2 \rightarrow e+C+CH+H^*$ | 26.27 | 26.2±0.5 | |
| | | $e+C_2H_2 \rightarrow e+C_2(d)+H+H^*$ | 27.47 | | |
| $H_\gamma$ (5 – 2) | 434.5 | $e+C_2H_2 \rightarrow e+C_2H+H^*$ | 17.93 | 20.0±1 | |
| | | $e+C_2H_2 \rightarrow e+C_2H(A)+H^*$ | 18.39 | | |
| | | $e+C_2H_2 \rightarrow e+C_2H(B)+H^*$ | 22.79 | 23.8±1 | |
| | | $e+C_2H_2 \rightarrow e+C_2+H+H^*$ | 23.93 | | |
| | | $e+C_2H_2 \rightarrow e+C+CH+H^*$ | 26.58 | 26.6±1 | |
| | | $e+C_2H_2 \rightarrow e+C_2H^++H^*$<br>$e+C_2H_2 \rightarrow e+C+CH(A)+H^*$<br>$e+C_2H_2 \rightarrow e+2C+H+H^*$ | 29.54<br>29.63<br>30.11 | 29.9±1 | |
| $H_\delta$ (6 - 2) | 410.5 | $e+C_2H_2 \rightarrow e+C_2H+H^*$<br>$e+C_2H_2 \rightarrow e+C_2H(B')+H^*$ | 18.10<br>21.75 | 22.0±1 | |
| | | $e+C_2H_2 \rightarrow e+C+CH+H^*$<br>$e+C_2H_2 \rightarrow e+C_2(d)+H+H^*$<br>$e+C_2H_2 \rightarrow e+C_2(C)+H+H^*$ | 26.74<br>27.94<br>28.46 | 28.8±1 | |
| | | $e+C_2H_2 \rightarrow e+C_2H^++H^*$<br>$e+C_2H_2 \rightarrow e+2C+H+H^*$ | 29.71<br>30.27 | 30.0±1 | |
| $H_\varepsilon$ (7 - 2) | 397.5 | $e+C_2H_2 \rightarrow e+C_2H+H^*$<br>$e+C_2H_2 \rightarrow e+C_2H(A)+H^*$ | 18.20<br>18.65 | 17.7±1 | |
| | | $e+C_2H_2 \rightarrow e+C_2+H+H^*$<br>$e+C_2H_2 \rightarrow e+C_2H(C)+H^*$ | 24.20<br>24.58 | 25.6±1 | |
| | | $e+C_2H_2 \rightarrow e+C+CH+H^*$<br>$e+C_2H_2 \rightarrow e+C_2(d)+H+H^*$<br>$e+C_2H_2 \rightarrow e+C_2(C)+H+H^*$ | 26.84<br>28.04<br>28.56 | 28.4±1 | |
| $H_\zeta$ (8 - 2) | 389.55 | $e+C_2H_2 \rightarrow e+C_2H+H^*$ | 18.26 | 21.1±0.5 | |
| | | $e+C_2H_2 \rightarrow e+C+CH+H^*$<br>$e+C_2H_2 \rightarrow e+C_2(d)+H+H^*$ | 26.91<br>28.10 | 28.0±0.5 | |
| C I (**$2s^22p3s$ $^1P^0$ - $2s^22p^2$ $^1D$**) | 193.1 | $e+C_2H_2 \rightarrow e+H_2+C+C^*$<br>$e+C_2H_2 \rightarrow e+CH+H+C^*$ | 20.20<br>21.19 | 22.0±1 | |
| | | $e+C_2H_2 \rightarrow e+2H+C+C^*$ | 24.72 | | |
| | | $e+C_2H_2 \rightarrow 2e+CH^++H(n=3,$<br>$4)+C^*$<br>$e+C_2H_2 \rightarrow e+H_2(u^3\Pi_u)+C+C^*$<br>$e+C_2H_2 \rightarrow 2e+H_2^++2C^*(^1P^o)$ | 42.95<br>43.61<br>43.21<br>43.33 | 42.4±1 | |
| C I (**2s$^2$2p3s $^1$P$^0$ – 2s$^2$2p$^2$ $^1$S**) | 248.4 | $e+C_2H_2 \rightarrow e+CH_2+C^*$<br>$e+C_2H_2 \rightarrow e+H_2+C+C^*$ | 18.21<br>20.20 | 19.6±1 | |
| | | $e+C_2H_2 \rightarrow e+CH+H+C^*$<br>$e+C_2H_2 \rightarrow e+2H+C+C^*$ | 21.19<br>24.72 | 24.2±1 | |
| | | $e+C_2H_2 \rightarrow 2e+CH^++H(n=3,$<br>$4)+C^*$<br>$e+C_2H_2 \rightarrow e+H_2(u^3\Pi_u)+C+C^*$<br>$e+C_2H_2 \rightarrow 2e+H_2^++2CI(^1P^o)$ | 42.95<br>43.61<br>43.21<br>43.33 | 42.2±1 | |
| C II (**2s$^2$4s $^2$S – 2s$^2$3p $^2$P$^0$**) | 391.8 | $e+C_2H_2 \rightarrow e+H_2+C_2^*$ | 36.70 | 35.9±0.5 | |
| | | $e+C_2H_2 \rightarrow e+H_2+C+C^*$ | 40.12 | 41.4±0.5 | |
| | | $e+C_2H_2 \rightarrow e+CH+H+C^*$ | 41.11 | | |
| | | $e+C_2H_2 \rightarrow e+2H+C+C^*$ | 44.64 | 44.6±0.5 | |
| | | $e+C_2H_2 \rightarrow 2e+CH_2^++C^*$ | 47.10 | 48.5±0.5 | |
| | | $e+C_2H_2 \rightarrow e+H_2+CIII+C^*$<br>$e+C_2H_2 \rightarrow$<br>$3e+2H^++CI(^1P^0)+C^*$ | 75.77<br>79.53 | 79.3±0.5 | |

Tab. 3. Calculated (based on enthalpy of formation) and experimentally determined threshold energies for $C_2$, CH, $CH^+$ and $C_2H$ and their comparison with previously published values. Dissociative channels for thresholds are proposed.

| Transition | Wavelength [nm] | Reaction | Calculated threshold energy [eV] | Measured threshold energy [eV] | Published value [eV] |
|---|---|---|---|---|---|
| $C_2$ $(d^3\Pi_g \rightarrow a^3\Pi_u)$ (6,5) | 468.7 | $e+C_2H_2 \rightarrow e+H_2+C_2^*$ | 10.18 | 11.00±0.5 | |
| | | $e+C_2H_2 \rightarrow e+2H+C_2^*$ | 14.70 | 15.7±0.5 | |
| | | $e+C_2H_2 \rightarrow e+H+H(n = 3-8)+C_2^*$ | 26.80 - 28.15 | 29.8±1 | |
| | | $e+C_2H_2 \rightarrow 2e+H+H^++C_2^*$ | 28.30 | | |
| $C_2$ $(C^1\Pi_g \rightarrow A^1\Pi_u)$ (1,0) | 361.2 | $e+C_2H_2 \rightarrow e+H_2+C_2^*$ | 10.92 | 11.1±0.5 | |
| | | $e+C_2H_2 \rightarrow e+2H+C_2^*$ | 15.44 | 16.2±0.5 | |
| | | $e+C_2H_2 \rightarrow e+H+H(n = 3-8)+C_2^*$ | 27.55 - 28.90 | 28.8±0.5 | |
| | | $e+C_2H_2 \rightarrow 2e+H+H^++C_2^*$ | 29.04 | | |
| $C_2$ $(C^1\Pi_g \rightarrow A^1\Pi_u)$ (2,1) | 359.9 | $e+C_2H_2 \rightarrow e+H_2+C_2^*$ | 11.14 | 11.2±0.5 | |
| | | $e+C_2H_2 \rightarrow e+2H+C_2^*$ | 15.66 | 16.0±0.5 | |
| | | $e+C_2H_2 \rightarrow e+H+H(n = 3-8)+C_2^*$ | 27.76 - 29.11 | 31.5±0.5 | |
| | | $e+C_2H_2 \rightarrow 2e+H+H^++C_2^*$ | 29.26 | | |
| $C_2$ $(D^1\Sigma_u^+ \rightarrow X^1\Sigma_g^+)$ | 232.3 | $e+C_2H_2 \rightarrow e+2H+C_2^*$ | 16.22 | 17.3±1 | |
| | | $e+C_2H_2 \rightarrow e+H+H(n = 3-8)+C_2^*$ | 28.32 - 29.67 | 33.0±1 | |
| | | $e+C_2H_2 \rightarrow 2e+H+H^++C_2^*$ | 29.82 | | |
| CH ($A^2\Delta$-$X^2\Pi$) (0.0) | 431.5 | $e+C_2H_2 \rightarrow e+CH+CH^*$<br>$e+C_2H_2 \rightarrow e+C+H+CH^*$<br>$e+C_2H_2 \rightarrow 2e+C^++H+CH^*$<br>$e+C_2H_2 \rightarrow 3e+C^++H^++CH^*$ | 13.02<br>16.55<br>27.81<br>41.41 | 13.5±0.5<br>17.0±0.5<br>27.8±0.5<br>41.2±0.5 | 13.5±0.5 (Beenaker & de Heer 1974) |
| CH ($B^2\Sigma^-$-$X^2\Pi$) (0.0) | 387.8 | $e+C_2H_2 \rightarrow e+CH+CH^*$<br>$e+C_2H_2 \rightarrow e+C+H+CH^*$<br>$e+C_2H_2 \rightarrow 2e+C^++H+CH^*$ | 13.34<br>16.87<br>28.13 | 13.8±0.5<br>16.9±0.5<br>28.5±0.5 | |
| CH ($C^2\Sigma^-$-$X^2\Pi$) (1.0) | 364.6 | $e+C_2H_2 \rightarrow e+CH+CH^*$<br>$e+C_2H_2 \rightarrow e+C+H+CH^*$<br>$e+C_2H_2 \rightarrow 2e+C+H^++CH^*$ | 13.56<br>17.09<br>30.69 | 14.3±0.5<br>18.1±0.5<br>30.9±1 | |
| $CH^+$ ($B^1\Delta$-$A^1\Pi$) (0.0) | 350.7 | $e+C_2H_2 \rightarrow 2e+CH+CH^{+*}$<br>$e+C_2H_2 \rightarrow 2e+C+H+CH^{+*}$ | 27.63<br>31.16 | 28.7±0.5<br>31.8±0.5 | |
| $C_2H$ continuum | 526 | $e+C_2H_2 \rightarrow H+C_2H^*$ | 7.21 | 9.7±0.5 | 9.5±0.2 (Okabe 1975) |

For convenience the values of the cross sections at several different energies for all the measured transitions are summarized in the table 4.

Tab. 4. Absolute values of the cross sections for all observed processes at the electron energies 30 eV and 50 eV.

| Transition | Wavelength [nm] | DEE CS at **25 eV** ($\times 10^{-19}$ $cm^2$) | DEE CS at **30 eV** ($\times 10^{-19}$ $cm^2$) | DEE CS at **40 eV** ($\times 10^{-19}$ $cm^2$) | DEE CS at **50 eV** ($\times 10^{-19}$ $cm^2$) | DEE CS at **60 eV** ($\times 10^{-19}$ $cm^2$) |
|---|---|---|---|---|---|---|
| $H_\alpha$ | 657 | 0.89 | 2.35 | 7.01 | 10.82 | 13.04 |
| $H_\beta$ | 486.7 | 0.28 | 1.01 | 2.59 | 4.64 | 5.64 |
| $H_\gamma$ | 434.5 | 0.08 | 0.58 | 0.96 | 1.49 | 1.37 |
| CH(A-X) | 431.5 | 10.61 | 10.70 | 15.56 | 20.20 | 20.87 |
| CH (B-X)(0,0) | 387.8 | 3.85 | 4.30 | 6.13 | 7.10 | 7.28 |
| CH (B-X)(1,0) | 364.6 | 0.93 | 1.24 | 4.08 | 6.18 | 7.58 |
| $CH^+$(B-A) | 350.7 | | 2.45 | 3.93 | 4.73 | 5.06 |
| $C_2$ (d-a)(6,5) | 468.7 | 7.41 | 8.80 | 10.32 | 12.51 | 13.52 |
| $C_2$ (C-A)(1,0) | 361.2 | 3.30 | 3.49 | 4.78 | 5.37 | 5.79 |
| $C_2$ (C-A)(2,1) | 359.9 | 2.33 | 2.56 | 3.51 | 4.10 | 4.38 |
| $C_2$(D-X) | 232.3 | 1.47 | 1.50 | 1.75 | 1.97 | 1.94 |
| C I (**$2s^22p3s$ $^1P^0$** | 193.1 | 0.56 | 1.71 | 2.73 | 4.19 | 5.43 |
| - $2s^22p2$ $^1D$) | | | | | | |
| C I ($2s^22p3s$ $^1P^0$ - $2s^22p^2$ $^1S$) | 248.4 | 1.30 | 2.33 | 3.66 | 6.01 | 7.58 |
| C II | 391.8 | | | 1.46 | 2.11 | 2.26 |
| $C_2H$ | 526 | 0.62 | 0.68 | 0.77 | 0.8 | |

Tab. 5. Absolute values of the cross sections for selected observed processes at selected electron energies E.

| H (3 – 2) | | H (4 – 2) | | CH ($A^2\Delta$-$X^2\Pi$) (0.0) | | $C_2H$ ($A^2\Pi$ - $X^2\Sigma^+$) | |
|---|---|---|---|---|---|---|---|
| E (eV) | DEE CS ($x10^{-19}$ $cm^2$) | E (eV) | DEE CS ($x10^{-19}$ $cm^2$) | E (eV) | DEE CS ($x10^{-19}$ $cm^2$) | E (eV) | DEE CS ($x10^{-19}$ $cm^2$) |
| 21 | 0.09 | 19.4 | 0.04 | 14 | 0.97 | 10 | 0.06 |
| 21.2 | 0.11 | 19.6 | 0.06 | 14.3 | 1.29 | 10.2 | 0.08 |
| 21.5 | 0.14 | 19.8 | 0.07 | 14.6 | 1.70 | 10.4 | 0.10 |
| 21.7 | 0.16 | 20 | 0.08 | 15 | 2.36 | 10.6 | 0.12 |
| 22 | 0.20 | 20.5 | 0.09 | 15.5 | 3.34 | 10.8 | 0.14 |
| 22.5 | 0.31 | 21 | 0.11 | 16 | 4.38 | 11 | 0.17 |
| 23 | 0.44 | 22 | 0.15 | 16.5 | 5.40 | 11.5 | 0.23 |
| 23.5 | 0.53 | 23 | 0.18 | 17 | 6.31 | 12 | 0.28 |
| 24 | 0.66 | 24 | 0.23 | 18 | 7.56 | 13 | 0.39 |
| 24.5 | 0.79 | 25 | 0.28 | 19 | 8.39 | 13.5 | 0.43 |
| 25 | 0.89 | 30 | 1.01 | 20 | 9.13 | 14 | 0.46 |
| 27 | 1.48 | 40 | 2.59 | 24 | 10.68 | 14.5 | 0.47 |
| 30 | 2.35 | 50 | 4.64 | 26.6 | 10.53 | 20 | 0.56 |
| 35 | 4.30 | 60 | 5.64 | 28 | 10.61 | 30 | 0.68 |
| 37 | 5.25 | 70 | 6.21 | 40 | 15.32 | 35 | 0.75 |
| 40 | 7.01 | 80 | 6.31 | 50 | 20.20 | 40 | 0.77 |
| 45 | 8.60 | 90 | 6.32 | 55 | 21.40 | 50 | 0.8 |
| 50 | 10.82 | 100 | 6.33 | 70 | 21.34 | | |
| 60 | 13.04 | | | 85 | 21.24 | | |
| 70 | 12.90 | | | 100 | 20.78 | | |

Conclusion

The dissociative excitation of the acetylene molecule upon the electron impact was studied experimentally using a crossed-beams apparatus and equipped with electron monochromator. The fluorescence spectrum of acetylene in UV/VIS spectral range was recorded and analysed. We have determined the absolute values of excitation-emission cross sections for particular excited fragments of acetylene and the threshold energies for individual fragments (C I, C II, H I (Balmer series) and molecular fragments CH I, CH II, $C_2H$ I, $C_2$ I). We have paid special attention to the continuum radiation of the ethynyl radical $C_2H$. Most of the determined thresholds and cross section values were never published before. On the basis of the comparison of the thermochemical data and present experimental data dissociative reactions active and contributing to the formation of these excited states were proposed.

Acknowledgements

This work was supported by the Slovak Research and Development Agency, project Nr. APVV-0733-11 and VEGA 1/0417/15.

This work has been carried out within the framework of the EUROfusion Consortium and has received funding from the Euratom research and training programme 2014-2018 under grant agreement No 633053. The views and opinions expressed herein do not necessarily reflect those of the European Commission.

Zubek, M. 1994, *JPhB*, 27, 573.

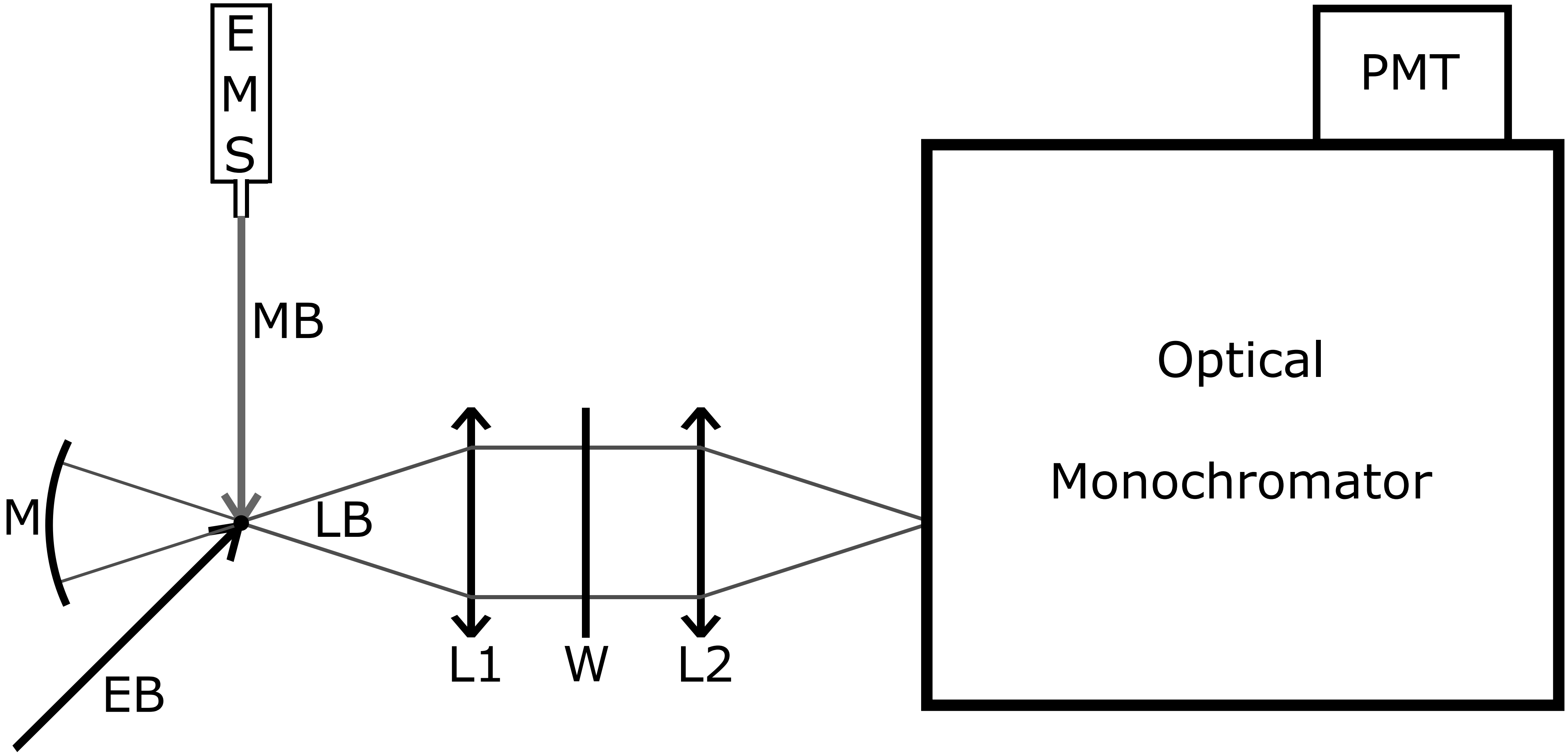

Fig. 1.

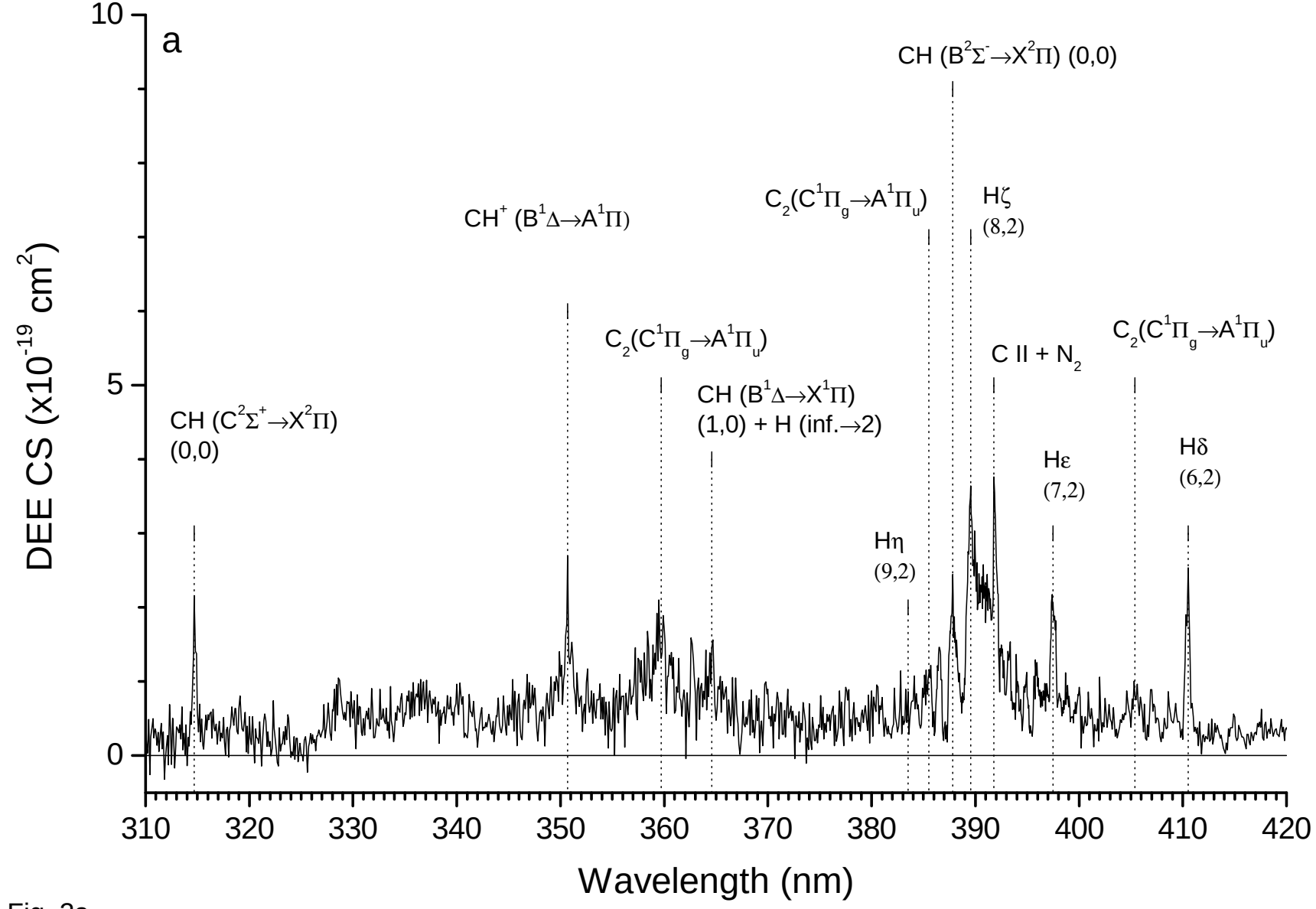

a
DEE CS (x10$^{-19}$ cm$^2$)
Wavelength (nm)
CH ($B^2\Sigma^- \rightarrow X^2\Pi$) (0,0)
$C_2(C^1\Pi_g \rightarrow A^1\Pi_u)$
Hζ
(8,2)
$CH^+$ ($B^1\Delta \rightarrow A^1\Pi$)
$C_2(C^1\Pi_g \rightarrow A^1\Pi_u)$
C II + $N_2$
$C_2(C^1\Pi_g \rightarrow A^1\Pi_u)$
CH ($B^1\Delta \rightarrow X^1\Pi$)
(1,0) + H (inf.→2)
CH ($C^2\Sigma^+ \rightarrow X^2\Pi$)
(0,0)
Hε
(7,2)
Hδ
(6,2)
Hη
(9,2)

Fig. 2a

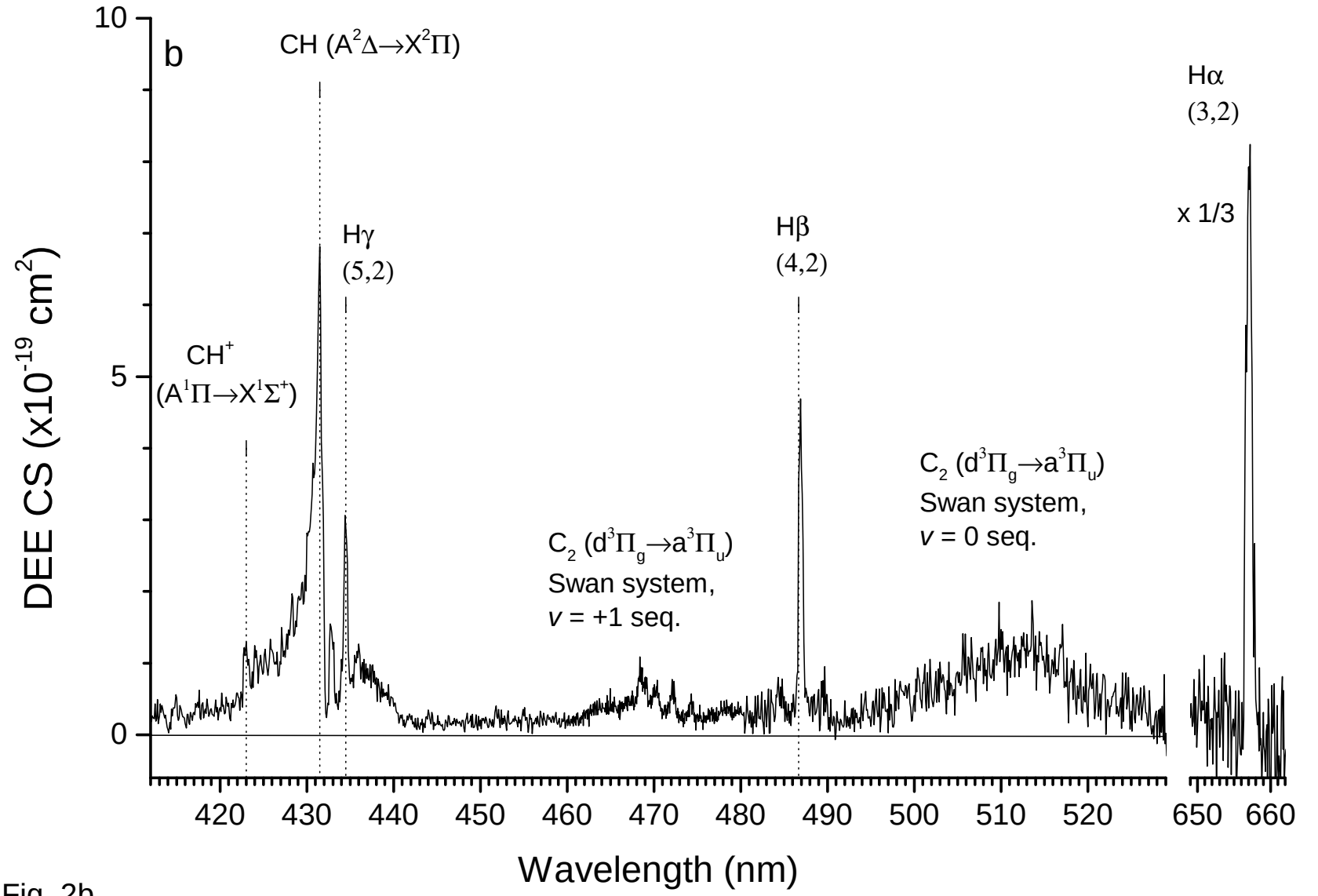

10
5
0
b
CH (A²Δ→X²Π)
Hγ
(5,2)
CH⁺
(A¹Π→X¹Σ⁺)
Hβ
(4,2)
Hα
(3,2)
x 1/3
C₂ (d³Πg→a³Πu)
Swan system,
v = +1 seq.
C₂ (d³Πg→a³Πu)
Swan system,
v = 0 seq.
420 430 440 450 460 470 480 490 500 510 520 650 660
DEE CS (x10⁻¹⁹ cm²)
Wavelength (nm)

Fig. 2b.

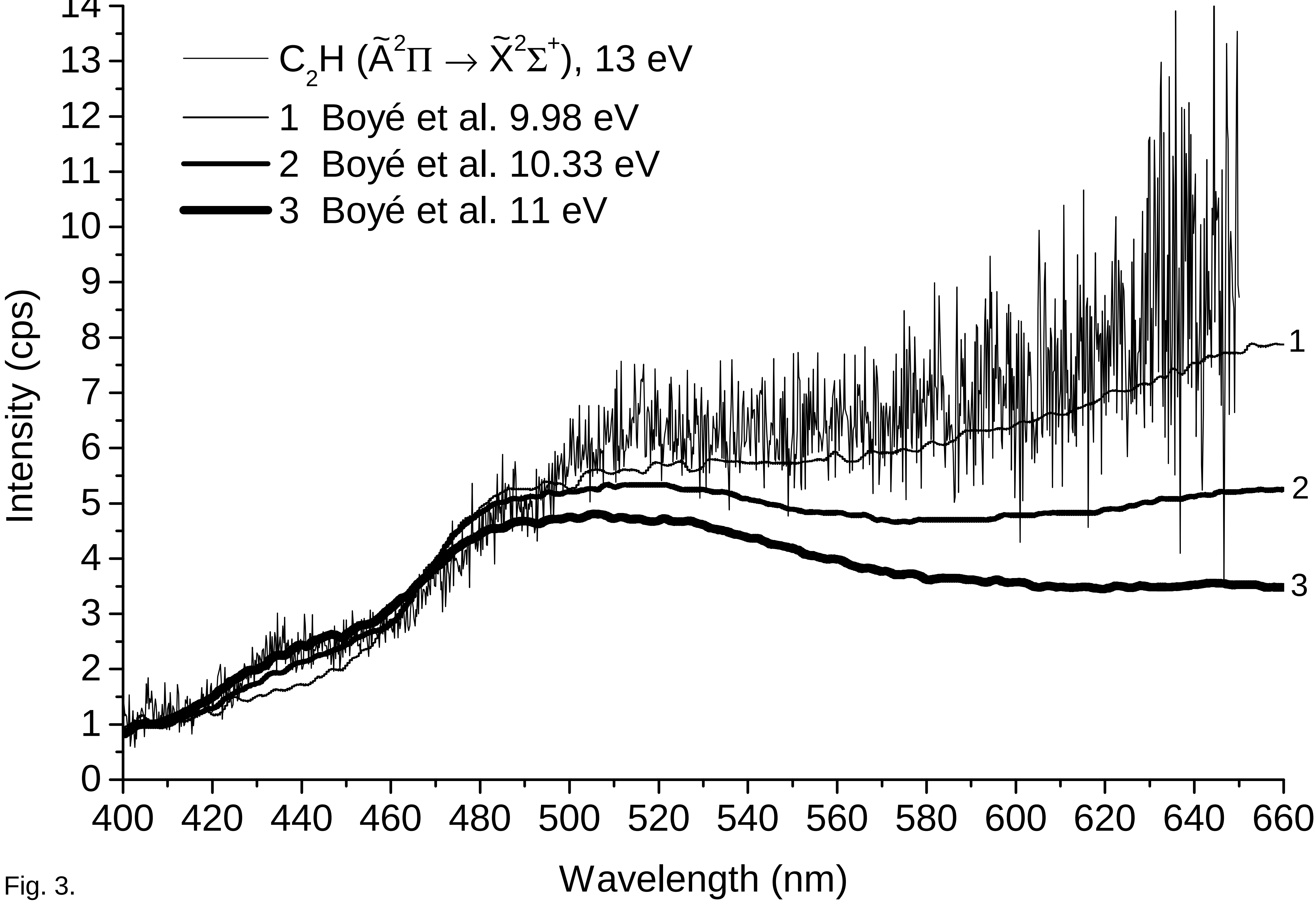

$C_2H$ ($\tilde{A}^2\Pi \rightarrow \tilde{X}^2\Sigma^+$), 13 eV
1 Boyé et al. 9.98 eV
2 Boyé et al. 10.33 eV
3 Boyé et al. 11 eV
Intensity (cps)
Wavelength (nm)
0
1
2
3
4
5
6
7
8
9
10
11
12
13
14
400
420
440
460
480
500
520
540
560
580
600
620
640
660
1
2
3

Fig. 3.

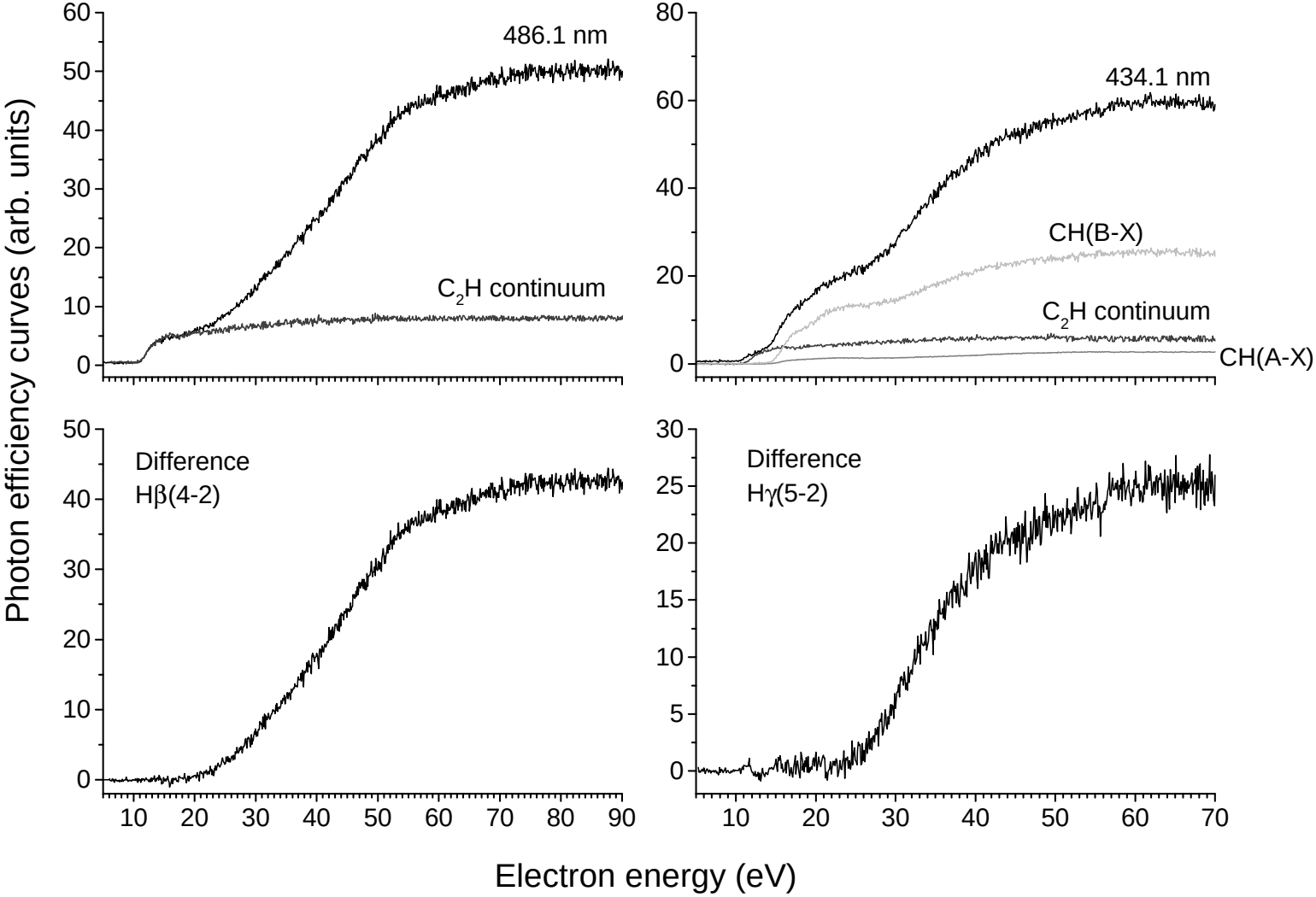

Fig. 4.

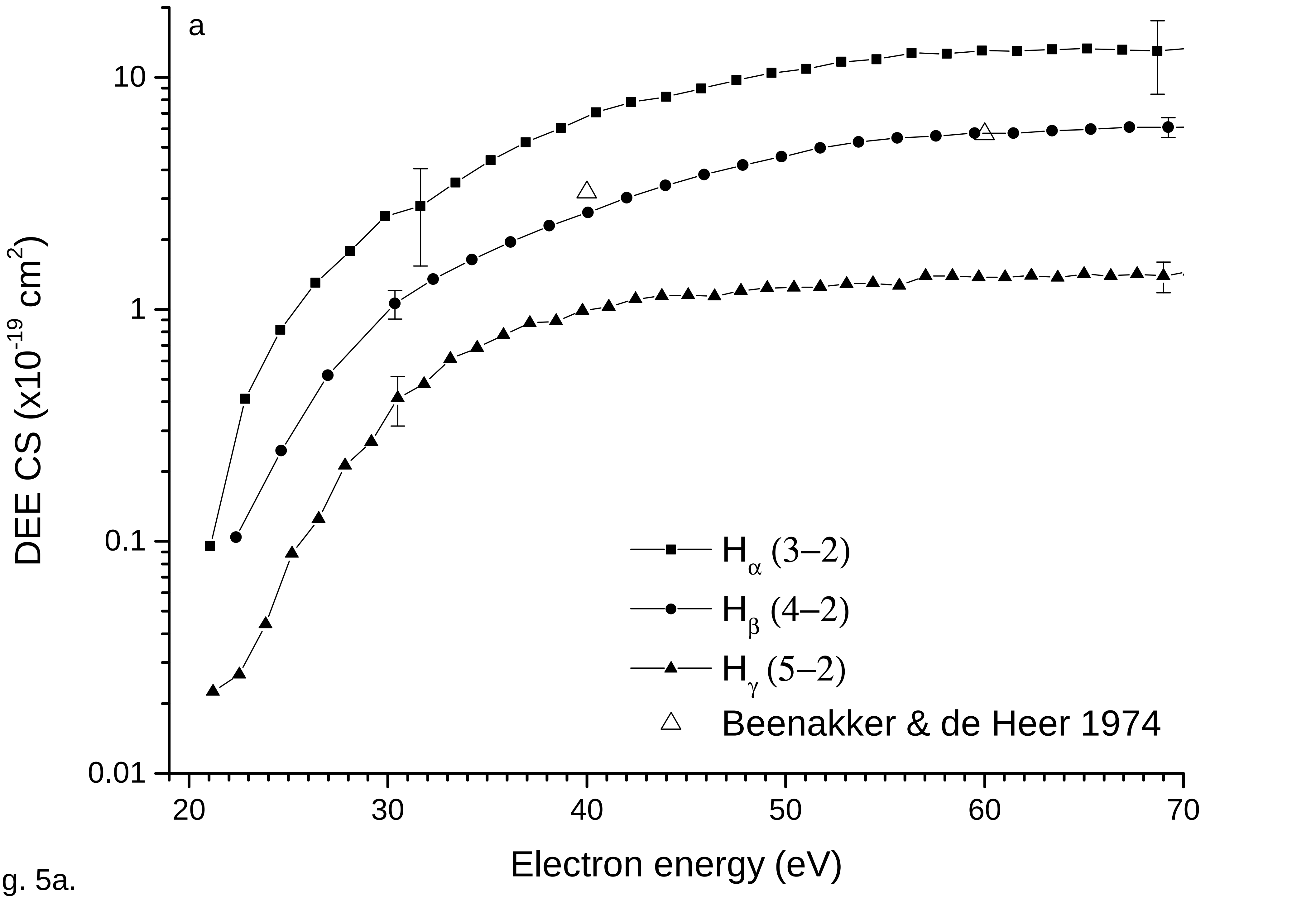

Fig. 5a.

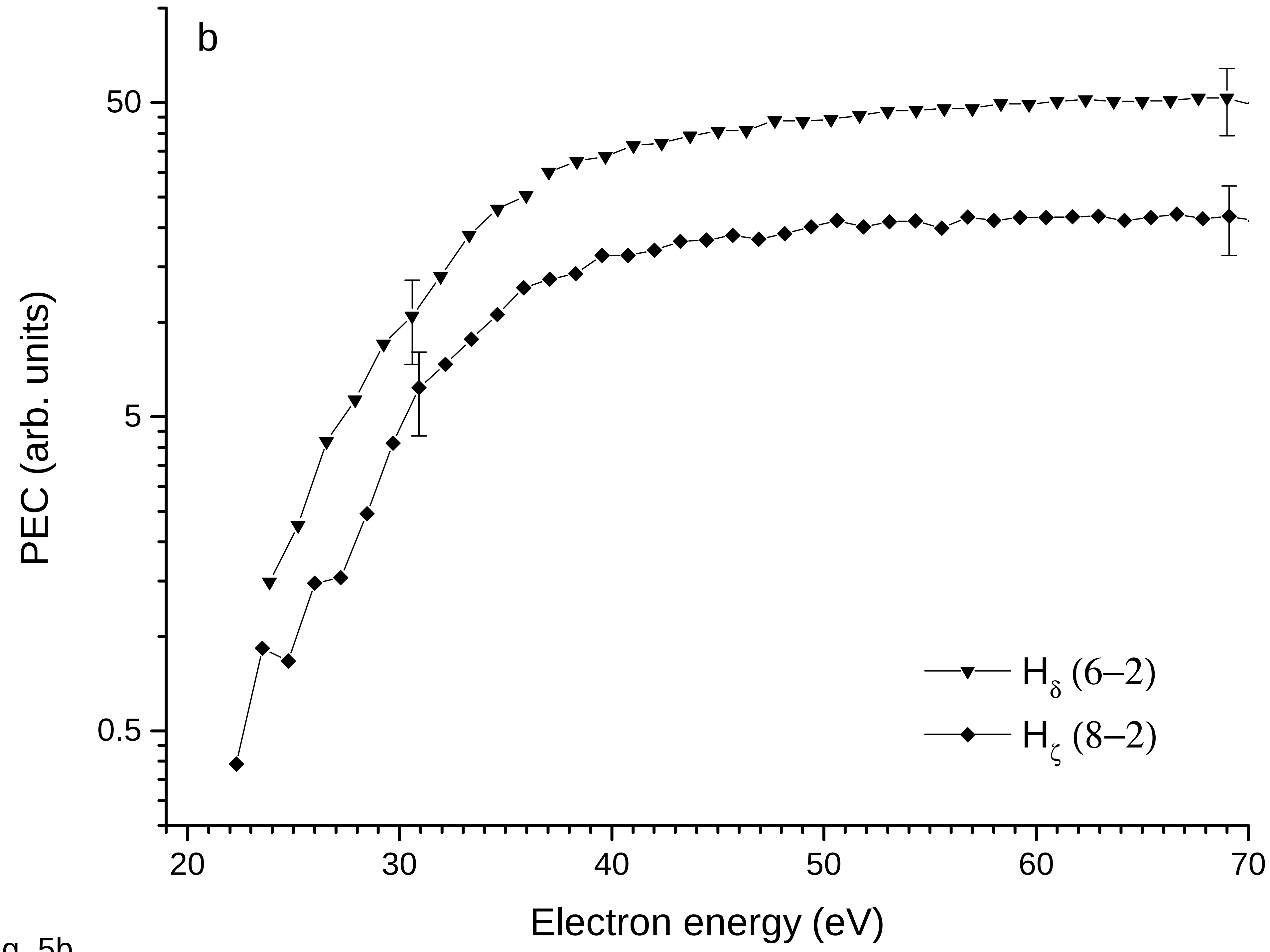

Fig. 5b.

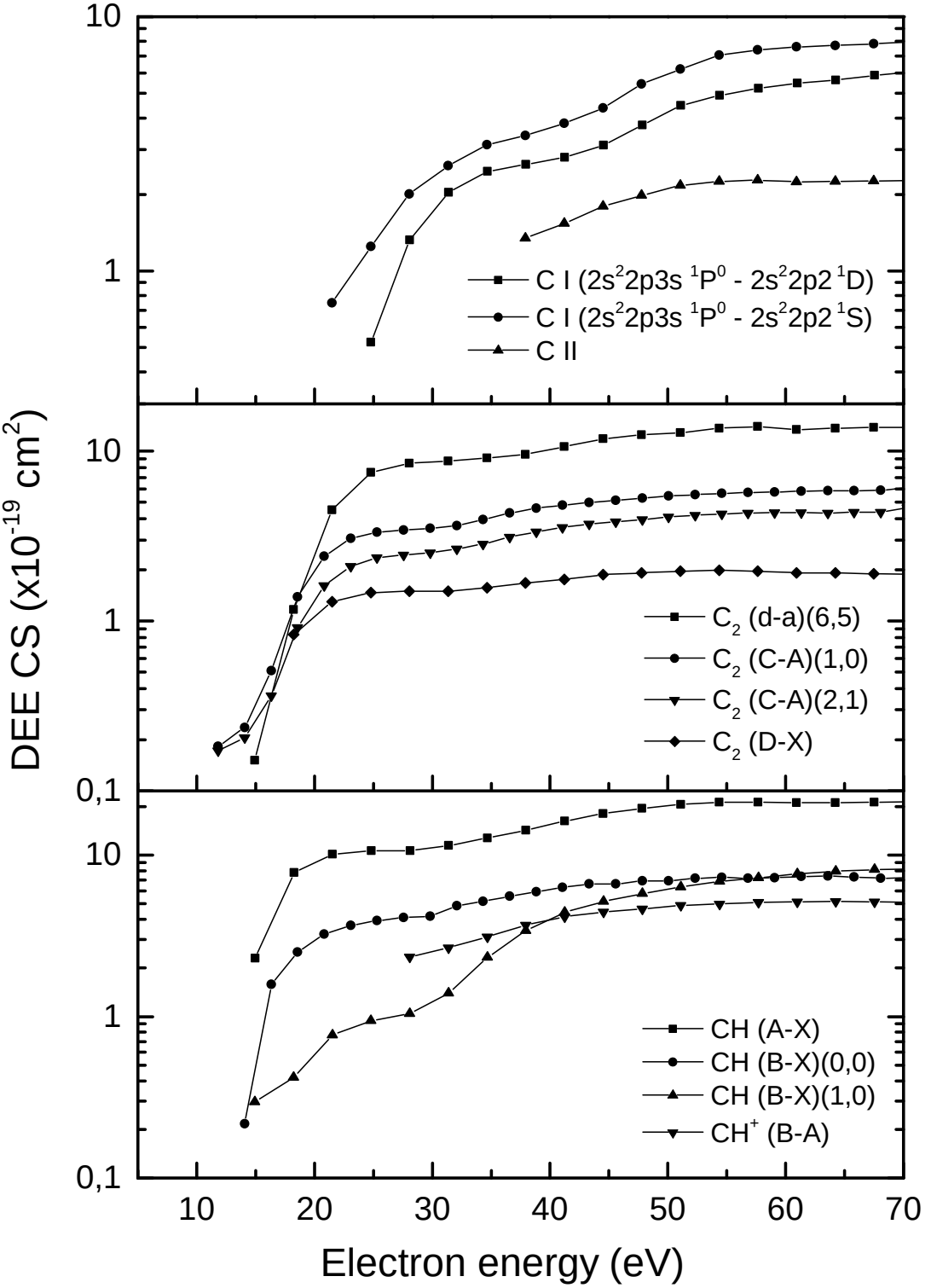

Fig. 6.

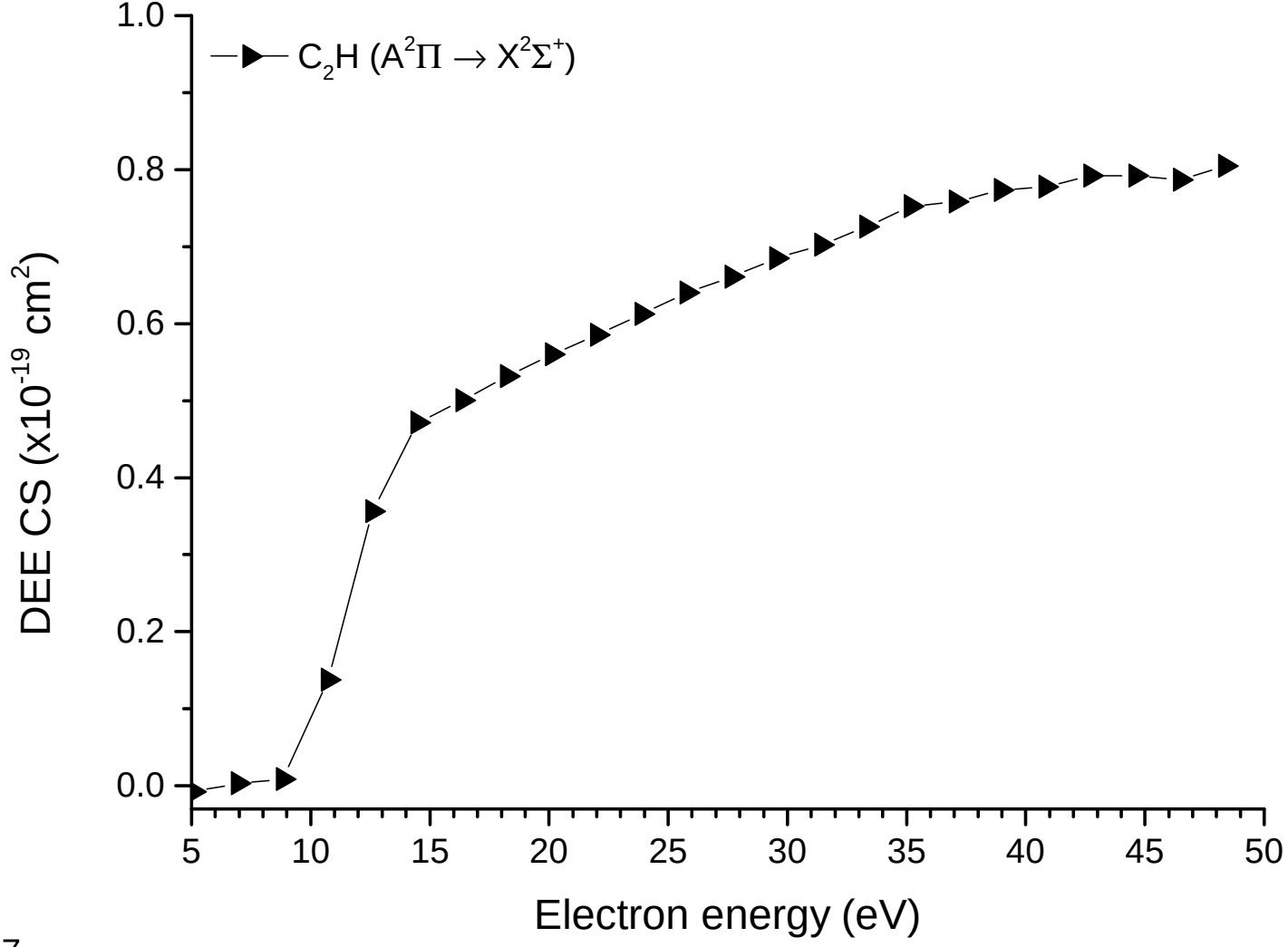

Fig. 7.

**Fig. 1.** The scheme of the experimental setup. EMS – effusive molecular source, MB – molecular beam, EB – electron beam, M – spherical mirror, L1, L2 – plano-convex fused silica lenses, W – $MgF_2$ window mounted in the wall of the vacuum chamber, PMT – photomultiplier tube detector.

**Fig. 2.** Acetylene spectrum in wavelength region of 310-420 nm (fig. 2a.) and 420-660 nm (fig. 2b.) at electron energy 50 eV with slits set to 100 μm (optical resolution ~0.4 nm).

**Fig. 3.** The continuum radiation of $C_2H$ fragment. The measured spectrum compared to data published by (Boyé et al. 2002) for different energies.

**Fig. 4.** Example of subtraction of the superposed cross section. Left – subtraction of $C_2H$ continuum signal from H(4 - 2) transition; right – subtraction of $C_2H$ continuum and CH($A^2\Delta$ - $X^2\Pi$) signals from H(5 - 2) transition.

**Fig. 5.** Cross sections for dissociative excitation of $H_2$ leading to emission of Balmer α, β and γ lines (fig. 5a.). Apparent Photon excitation curves – relative excitation functions of Balmer δ and ζ which suffer from the incomplete detection problem (fig. 5b.). It was impossible to determine the $H_\varepsilon$(7 - 2) cross section due to strong overlap with the $H_2$ bands.

**Fig. 6.** Cross sections for dissociative excitation processes of $C_2H_2$ resulting in emission of C, $C_2$ and CH fragments.

**Fig. 7.** PEC for $C_2H$ continuum calibrated to cross section values measured at 526 nm.